# Orientation Echoes via Concerted Terahertz and Near-IR excitation

Ran Damari[1,2], Amit Beer[1,2], Sharly Fleischer[1,2,*]

[1]Raymond and Beverly Sakler Faculty of Exact Sciences, School of Chemistry, Tel Aviv University, Tel Aviv 6997801, Israel
[2]Tel-Aviv University center for Light-Matter-Interaction, Tel Aviv 6997801, Israel

Abstract:

A new and efficient method for orientation echo spectroscopy is presented and realized experimentally. The excitation scheme utilizes concerted rotational excitations by both ultrashort terahertz and near-IR pulses and its all-optical detection is enabled by Molecular Orientation Induced Second Harmonic method [*J.Phys.Chem.A 126, 3732-3738 (2022)*]. The method provides practical means for orientation echo spectroscopy of gas phase molecules and highlights the intriguing underlying physics of coherent rotational dynamics induced by judiciously-orchestrated interactions with both resonant (terahertz) and nonresonant (NIR) light pulses.

**Section I: Introduction**

Laser-controlled rotational dynamics emerged from rotational coherence spectroscopy [1,2], a field aimed to extract rotational coefficients and structural data of molecules by monitoring their light-induced rotational dynamics. While primarily developed to induce anisotropic angular distributions of molecules within, otherwise isotropic molecular ensembles, the field of laser-controlled rotational dynamics has evolved into a plethora of tangential scientific applications in the realm of "molecular frame spectroscopy" and has yielded various intriguing observations and new phenomena over the years (for review papers see [3–7]). Rotational echo spectroscopy [8–17], selective rotation of molecular enantiomers [18–21] and advanced rotational imaging [22,23] techniques are merely few recent achievements of this vibrant field of research.

Anisotropic angular distributions are categorized as 'alignment' and 'orientation' with the inversion symmetry of the medium retained in the former or transiently lifted in the latter respectively. In recent years, vast efforts are put into obtaining molecular orientation that provides a more refined control over the angular distribution of molecules. This requires direct controllability over the molecular dipoles using a DC-or quasi DC fields [24–28], two-color ($\omega + 2\omega$) optical schemes [29], or via intense single-cycle terahertz pulses [30–35] used in this work. Here we demonstrate uniquely desirable rotational responses of polar gas molecules that are induced via judiciously-orchestrated rotational excitation by a terahertz (THz) and near-IR (NIR) pulses.

The paper is organized as follows: section II describes the experimental setup and the all-optical detection scheme. Section III presents the experimental results of the THz-delay-NIR excitation scheme and section IV presents the transition pathways that govern the echo signal.

## Section II: Experimental system and detection scheme

The experimental setup used in this work is shown in Fig. 1. A short NIR pulse (800nm, 100 fs) is split in three beams. The first beam is used to generate a single-cycle THz pulse ($E_{THz}$) via tilted pulse front THz generation in $LiNbO_3$ prism [30]. The other two beams are routed through computer-controlled delay stages and provide a NIR pump ($P_{NIR}$) and NIR probe ($P_{probe}$).

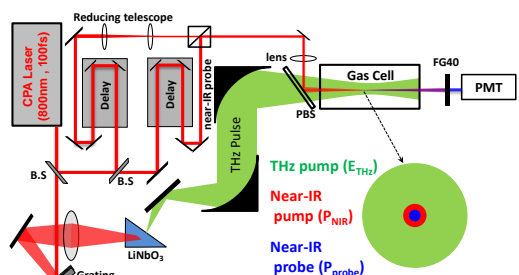

Figure 1: Experimental setup (described in the text). Note the cross section of the three beams at the focus in the gas, with the probe beam waist (blue) smaller than both THz (green) and NIR pump (red) beams to ensure monitoring the interaction region of the two. B.S – beam splitter, PBS – pellicle beam splitter, FG40 – short pass filter that transmits the 400nm signal and blocks the 800nm probe.

The diameter of the NIR pump is reduced to half its initial diameter (10mm→5mm) using a telescope and recombined with the NIR probe. The two NIR beams are focused by a lens into the gas cell at the center of the (~2mm) THz beam waist. This ensures that the waist of $P_{probe}$ is ~2 times smaller than that of $P_{NIR}$ such that the second harmonic signal (SH) generated by $P_{probe}$ emanates from the interaction volume that is affected by both the THz (green) and NIR (red) beams (see enlarged cross-section in Fig.1). The detection scheme relies on the recently published MOISH technique [36]. Briefly, MOISH relies on the inversion asymmetry of the gas medium upon the orientation of its polar molecular constituents. The latter gives rise to an effective nonlinear susceptibility $\chi^{(2)}$ that manifest as SH of the NIR (800nm) probe pulse and detected as a 400nm signal. We note that in general, the MOISH signal may be accompanied by an electronic contribution induced by the incident THz field, as in TFISH [37]. Moreover, the FID emission from the oriented gas at the quantum rotational revival time induces a secondary TFISH signal that interferes with the MOISH signal [36]. In addition, phase mismatch may further complicate the MOISH signal, owing to the long interaction length of the NIR probe with the THz field (with Rayleigh range of few centimeters for the latter). This invokes practical limitations on the gas density that can be used since the phase mismatch ($\Delta k$) increases with $L$ - the effective interaction length and $P$- the gas pressure (density) [36]. In the concerted excitation scheme presented hereafter, all of the above concerns are practically lifted since the orientation responses induced by the action of both $E_{THz}$ and $P_{NIR}$ do not overlap temporally with the incident THz field or its subsequent FID emissions at revival times, hence governed solely by the net contributions of molecular orientation. In addition, the effective length of interaction $L$ is significantly reduced to the mutual volume of $E_{THz}$ and $P_{NIR}$, practically dictated by the latter and enables phase-matched SHG at larger gas densities.

## Section III: Orientation echoes induced by THz – delay – NIR scheme

Echo spectroscopy emerged into the field of rotational dynamics only recently [10], more than six decades after its first realization in magnetic resonance [38] and many years after its implementation in electronic [39] and vibrational [40] spectroscopy. The basic echo scheme is generic to all the above; echo responses are induced by two pulses with controlled delay apart.

The 1st first pulse induces coherent dynamics that evolves during some 'waiting time' $\Delta t$ and the 2nd pulse effectively reverses the direction of phase accumulation and gives rise to a 'rephased echo signal' at t=2$\Delta t$. While the optimal echo sequence incorporates pulse areas of $\pi/2$ and $\pi$ for the 1st and 2nd pulses respectively, echo signals can be induced by pulses with smaller 'flip angles' [41]. Owing to their multi-level nature (with several tens of populated levels at ambient temperature and broad-band pulses covering multiple transition frequencies among these levels), molecular rotors present intriguing dynamical features emanating from interferences between multiple transition pathways within the rotational level manifold [15,17]. Orientation echoes in the gas phase have been realized using a THz-delay-THz sequence where the 1st THz pulse induces one quantum coherence terms (1QC) $|J\rangle\langle J| \to |J\rangle\langle J+1|$ via one dipole interaction and the 2nd THz pulse partially rephases these terms (1QC*) $|J\rangle\langle J+1| \to |J+1\rangle\langle J+1| \to |J+1\rangle\langle J|$ via two instantaneous dipole interactions [8,42].

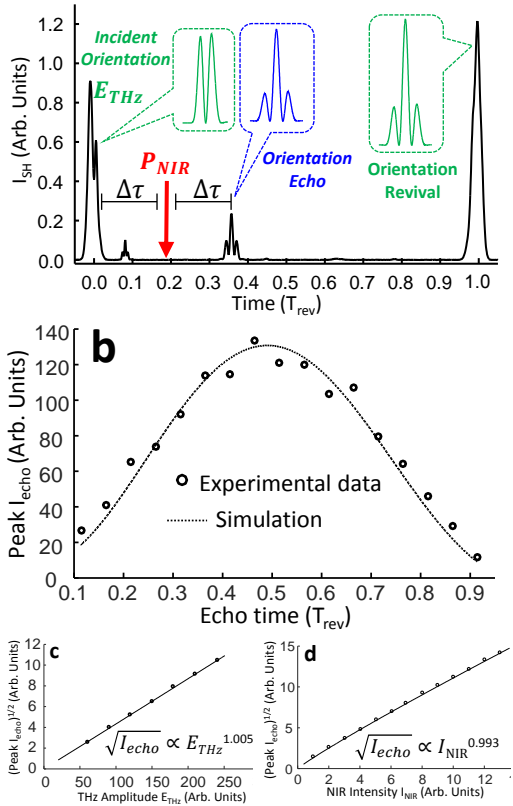

In what follows we show that THz-induced 1QCs can be efficiently rephased by a second, short NIR pulse that interacts with the molecular rotors via their anisotropic polarizability (rather than field-dipole interaction) and results in a rephasing orientation echo signal. Figure 2a depicts the experimental SH signal from $CH_3I$ gas (10 torr, room temperature) as a function of time (given in units of $T_{rev}$=66ps). A THz field ($E_{THz}$, t=0) is followed by a NIR pulse ($P_{NIR}$, t=0.18$T_{rev}$) and an orientation echo signal is detected at t=0.36$T_{rev}$ (i.e. at 2$\Delta t$). The SH signal at t=0 results from both the instantaneous electronic and orientation contributions to the nonlinear susceptibility $\chi^{(2)}$ [36]. Note that $P_{NIR}$ (t=0.18) does not induce any instantaneous orientation, hence lacks any incident SH signal. However, it induces rephasing rotational dynamics that manifest as an orientation echo signal at t=0.36$T_{rev}$. We wish to stress the excellent agreement between the temporal shape of the orientation echo signal and the simulated triple peak shape (blue inset), as opposed to the incident (t=0) and revival (t=1$T_{rev}$) signals that are significantly distorted with respect to the simulated transients (green insets). These distortions result from the interference between the THz-induced electronic and the nuclear

Figure 2: (a) Experimental THz-delay-NIR echo signal from 10torr $CH_3I$ gas at room temperature. A THz field applied at t=0 is followed by a NIR pulse at t=0.23$T_{rev}$ (=15.2ps) and induces an orientation echo signal at t=0.46$T_{rev}$. (b) depicts the peak of the orientation echo signal as a function of the delay between the THz and NIR. The maximal echo response is observed at $2\Delta t = 0.5 Trev$, i.e. for delay of 0.25$T_{rev}$ between the THz and NIR and with excellent agreement with the theoretical curve (dashed black). (c,d) shows the dependence of the peak echo signal intensity on the THz amplitude and NIR intensity, with linear dependence on both respectively.

(orientation) contributions to $\chi^{(2)}$ and due to severe phase-mismatch effects that are amplified by the long interaction length of $E_{THz}$ and $P_{probe}$ [36]. For the echo signal of interest in this work however, the effective interaction length of the concerted THz and NIR pulses (from which the echo signal emerges) is primarily dictated by the Rayleigh length of the NIR pump (~3mm) i.e. significantly shorter than that of the THz pulse (few cm). Hence the phase-matching ramifications as well as the Gouy phase-shift experienced by the focusing THz field are practically relieved and enable phase-matched detection of the orientation echo response. For the sharp-eyed readers we note that the additional signal observed at t=0.09Trev arises from "role-switching" of $P_{NIR}$ and $P_{probe}$, i.e. when the delay of $P_{probe}$ is ~half the delay of $P_{NIR}$ (that is set at 0.18$T_{rev}$ in Fig.2a), an orientation echo signal is also observed.

Figs.2c,d show the echo signal dependence on the THz field amplitude (Fig.2c) and the NIR pulse intensity (Fig.2d) respectively. Note that the homodyne MOISH signal is proportional to $\left|\chi^{(2)}\right|^2$, hence the linear dependence of the square-root of the echo ($\sqrt{I_{echo}}$) on both excitation pulses indicates that the echo response is induced by one field-dipole and one Raman interaction with the THz and NIR pulses respectively. This marks a key difference of the THz-delay-NIR scheme from the THz-delay-THz orientation echo schemes as the latter includes one THz-dipole and two THz-dipole interactions with the 1$^{st}$ and 2$^{nd}$ pulses respectively [8,15]. The latter relieves the need for a 2$^{nd}$ intense THz pulse and significantly increases the feasibility of the THz-delay-NIR scheme. For comparison, while ref. [8] used high THz-field amplitudes (~600 kV/cm) for studying acetonitrile gas (CH$_3$CN, 70torr) chosen for its large dipole moment ($\mu = 3.92 \, Debye$), here we obtain rotational echo signals from CH$_3$I ($\mu = 1.8 \, Debye$) at the single torr pressures. Figure 2b shows the peak of the echo signal as a function of delay ($\Delta t$) between the THz and NIR pulses with fixed amplitude/Intensity respectively. Each data point represents the peak echo signal obtained in an experiment just like the one shown in Fig.2a, only with varying delays between $E_{THz}$ and $P_{NIR}$. The maximal echo response is obtained at $\Delta t$=T$_{rev}$/4, in excellent agreement with the calculated results shown by the dashed black curve. We included the expected collisional decay for 10torr CH$_3$I that was measured previously by multiplying the simulated (decay-free) curve by $\exp[-\gamma t]$ where $\gamma = 6.4 \cdot 10^{-2} T_{rev}^{-1}$ (for collisional decay of CH$_3$I see [35]). The dependence of I$_{echo}$ on $\Delta t$ is closely reminiscent to that obtained in NIR-delay-NIR alignment echoes [15], only in the latter, the alignment echo response peaks at $\Delta t$=T$_{rev}$/8. In what follows we analyze and compare between the THz-delay-THz excitation scheme of [8] and the THz-delay-NIR scheme presented here, and highlight delicate, yet fundamental differences between the two.

*Section IV: Multiple Excitation Pathways in Orientation Echoes*

Figure 3 compares the excitation pathways that govern the two orientation-echo schemes discussed above. We utilize a graphical presentation inspired by the rotational density matrix formalism [15,43]. In order to simplify the discussion, we follow the evolution of the coherence terms induced on the right-hand-side of the diagonal and their transition pathway toward their rephasing counterpart (on the left side of the diagonal). Figure 3a presents the THz-delay-THz scheme [8] as a reference for the 'typical' echo scheme in multi-level rotational system. Here the 1$^{st}$ THz field interacts with population terms $|J\rangle\langle J|$ and $|J+1\rangle\langle J+1|$ via the molecular dipole, $\vec{\mu}\cdot\vec{E}$, to create the 1QC term $|J\rangle\langle J+1|$ (green arrows leading to 1QC term marked by the open circle). The latter evolves under the field-free Hamiltonian for time $\Delta t$, during which it accumulates a phase, $\Delta\varphi = -(E_{J+1} - E_J)\Delta t/\hbar$. Now comes the second THz field and interacts twice with the 1QC. The first interaction creates population terms $|J\rangle\langle J|$ and $|J+1\rangle\langle J+1|$ (dashed red arrows) and the second interaction creates the rephasing 1QC* term $|J+1\rangle\langle J|$ (dashed blue arrows). The 1QC* (stripped circle) evolves for another $\Delta t$ during which it accumulates the same phase, only with a negative sign $\Delta\varphi = -(E_J - E_{J+1})\Delta t/\hbar$, which results in the effective 'rephasing'. At t=2$\Delta t$, all of the 1QC terms regain the absolute phase they were 'born' with at the end of the first THz pulse, and an orientation echo signal is observed. Adjacent population terms that are discarded from the above description also contribute to the 1QC* amplitude as depicted by the dashed orange arrows, for example, the 1QC term of interest is partially projected onto the $|J\rangle\langle J-1|$ term that serves as the rephasing coherence of $|J-1\rangle\langle J|$. This manifests as a compromised degree of echo rephasability and by the dependence of the echo signal on $\Delta t$ (Fig.2d) and discussed above.

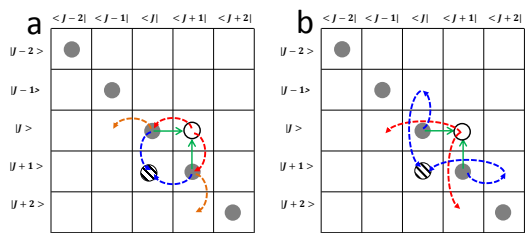

Figure 3: Graphical presentation of the interfering pathways for (a) THz-delay-THz scheme and (b) THz-delay-THz scheme. 'Short' arrows depict THz-induced transitions (via the dipole interaction term) and 'Long' arrows depict NIR-induced transition (via stimulated Raman).

Figure 3b depicts the excitation pathways that govern the THz-delay-NIR echo of interest in this work. The 1$^{st}$ THz field induces 1QC as in Fig.3a. The 2$^{nd}$ pulse however, is a NIR pulse that interacts only once via the anisotropic polarizability of the molecules. The latter induces $|J\rangle\langle J+1| \to |J\rangle\langle J-1|$ and $|J\rangle\langle J+1| \to |J+2\rangle\langle J+1|$ transitions (dashed red arrows), namely it creates rephasing 1QC* terms for the neighbouring 1QCs of $|J\rangle\langle J+1|$ but not for its original 1QC. In fact, the rephasing term for the 1QC $|J\rangle\langle J+1|$ induced by the THz field (i.e. the 1QC* term $|J+1\rangle\langle J|$) is contributed by $|J-1\rangle\langle J|$ and $|J+1\rangle\langle J+2|$ via one Raman transition with each. A simple analysis of the phase difference accumulated between these two 1QC terms serves to explain the time dependence of the echo signal on $\Delta t$ as revealed by Fig.2d; Once fully 'born' via the interaction with the THz field, each 1QC term accumulates a phase given by: $\Delta\varphi_{|J+1\rangle\langle J+2|} = (E_{J+2} - E_{J+1})\cdot\Delta t/\hbar = (2J+4)\pi\Delta t$ and $\Delta\varphi_{|J-1\rangle\langle J|} = (E_J - E_{J-1})\cdot\Delta t/\hbar = 2J\pi\Delta t$ where $\Delta t$ is given in $T_{rev}(=1/2B)$ units. The maximal echo response is observed when the phase difference between the two interfering 1QCs becomes $\Delta\Delta\varphi = 4\pi\Delta t = \pm\pi$, namely at $\Delta t = 1/4\,T_{rev}, 3/4\,T_{rev}, 5/4\,T_{rev}\ldots$ as shown in Fig.2b.

The transition pathways that govern the THz-delay-NIR echo are further revealed by considering the response of a single 1QC term (induced by the 1$^{st}$ THz field) to the 2$^{nd}$ NIR pulse. For this task, we simulated a narrow-band, multicycle THz field that couples only two rotational states $|J\rangle$ and $|J+1\rangle$ and creates only a single $|J\rangle\langle J+1|$ coherence term ($|5\rangle\langle 6|$ in Fig.4b).

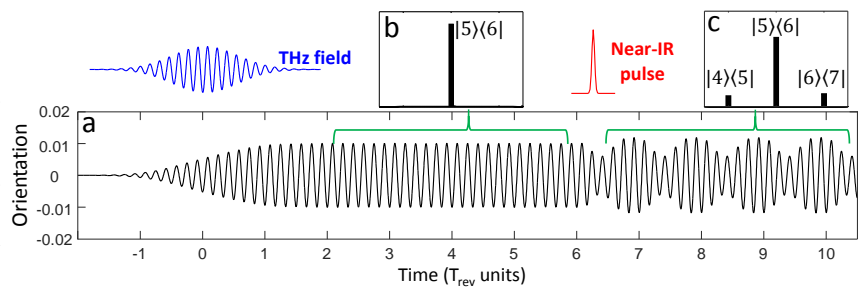

Figure 4: Simulated orientation dynamics induced by a multicycle THz field (blue) and a delayed ultrashort NIR pulse (red). (a) time domain orientation dynamics. (b) Spectral domain analysis of the dynamics between the two pulses and (c) following the NIR pulse.

Following its creation and field-free evolution ($2T_{rev} < t < 6T_{rev}$), a short NIR pulse is applied and results in periodic modulation of the orientation dynamics as observed in Fig.4a at $t>6T_{rev}$. Spectral analysis of the orientation dynamics at $t>6T_{rev}$ reveals two additional side peaks at frequencies $2.5 cm^{-1}$ and $3.5 cm^{-1}$ that correspond to the $|4\rangle\langle5|$ $and$ $|6\rangle\langle7|$ coherence terms respectively (Fig.4c). These are induced via one Raman transition from the $|5\rangle\langle6|$ term as predicted by the red dashed arrows in Fig.3b. We further note that in the excitation scheme of Fig.4, we found that the depth of modulation (which is directly associated with the rephasing coherences amplitudes) is independent of $\Delta t$, providing clear indication that the $\Delta t$ dependence of the echo signal stems from multiple pathway interferences discussed above. Hence, when only a single coherence term is induced by the THz field (as in Fig.4), $\Delta t$ merely governs the time at which the NIR pulse is applied but does not alter the rephasing coherence amplitude.

In conclusion, we have demonstrated an efficient, all-optical excitation scheme for inducing orientation echo signals in gas phase molecular rotors. The detection of the orientation echo signal relies on the SH signal enabled by the lifted inversion symmetry of the medium upon orientation (MOISH) and emanate from the mutual interaction volume of the THz and NIR excitation pulses. The presented method provides new means for orientation echo spectroscopy in a broad range of gas pressures and molecular entities. It should be noted that the presented technique is applicable to linear and symmetric top molecular rotors, where dipole and stimulated Raman interactions drive a mutual rotational degree of freedom. Concerted THz- and NIR- induced rotations provide uniquely intriguing possibilities in three dimensional dynamics and coherent rotational control of asymmetric molecular rotors.

Acknowledgments

The authors acknowledge the support of the Israel Science Foundation (926/18), the Wolfson family foundation (PR/ec/20419) and the PAZI foundation.

**References**

[1]   P. M. Felker, J. Phys. Chem. **96**, 7844 (1992).
[2]   P. M. Felker, J. S. Baskin, and A. H. Zewail, J. Phys. Chem **90**, 124 (1986).
[3]   H. Stapelfeldt and T. Seideman, Rev. Mod. Phys. **75**, 543 (2003).
[4]   M. Lemeshko, R. V. Krems, J. M. Doyle, and S. Kais, Mol. Phys. **918**, 1648 (2013).
[5]   Y. Ohshima and H. Hasegawa, Int. Rev. Phys. Chem. **29**, 619 (2010).
[6]   S. Fleischer, Y. Khodorkovsky, E. Gershnabel, Y. Prior, and I. S. Averbukh, Isr. J. Chem. **52**,

(2012).

- [7] C. P. Koch, M. Lemeshko, and D. Sugny, Rev. Mod. Phys. **91**, 035005 (2019).
- [8] J. Lu, Y. Zhang, H. Y. H. Y. H. Y. Hwang, B. K. B. K. Ofori-Okai, S. Fleischer, and K. A. K. A. K. A. Nelson, Proc. Natl. Acad. Sci. **113**, 11800 (2016).
- [9] D. Rosenberg, R. Damari, S. Kallush, and S. Fleischer, J. Phys. Chem. Lett. **8**, 5128 (2017).
- [10] G. Karras, E. Hertz, F. Billard, B. Lavorel, J.-M. Hartmann, O. Faucher, E. Gershnabel, Y. Prior, and I. S. Averbukh, Phys. Rev. Lett. **114**, 153601 (2015).
- [11] J. Ma, L. H. Coudert, F. Billard, M. Bournazel, B. Lavorel, J. Wu, G. Maroulis, J.-M. Hartmann, and O. Faucher, Phys. Rev. Res. **3**, 023192 (2021).
- [12] J.-M. Hartmann, J. Ma, T. Delahaye, F. Billard, E. Hertz, J. Wu, B. Lavorel, C. Boulet, and O. Faucher, (2020).
- [13] K. Lin, P. Lu, J. Ma, X. Gong, Q. Song, Q. Ji, W. Zhang, H. Zeng, J. Wu, G. Karras, G. Siour, J.-M. Hartmann, O. Faucher, E. Gershnabel, Y. Prior, and I. S. H. Averbukh, Phys. Rev. X **6**, 041056 (2016).
- [14] D. Rosenberg and S. Fleischer, Phys. Rev. Res. **2**, 023351 (2020).
- [15] D. Rosenberg, R. Damari, and S. Fleischer, Phys. Rev. Lett. **121**, 234101 (2018).
- [16] K. Lin, J. Ma, X. Gong, Q. Song, Q. Ji, W. Zhang, H. Li, P. Lu, H. Li, H. Zeng, J. Wu, J.-M. Hartmann, O. Faucher, E. Gershnabel, Y. Prior, and I. S. Averbukh, Opt. Express **25**, 24917 (2017).
- [17] Z. Lian, Z. Hu, H. Qi, D. Fei, S. Luo, Z. Chen, and C.-C. Shu, Phys. Rev. A **104**, 053105 (2021).
- [18] I. Tutunnikov, J. Floß, E. Gershnabel, P. Brumer, I. S. Averbukh, A. A. Milner, and V. Milner, Phys. Rev. A **101**, 021403 (2020).
- [19] M. Leibscher, T. F. Giesen, and C. P. Koch, J. Chem. Phys. **151**, 014302 (2019).
- [20] I. Tutunnikov, E. Gershnabel, S. Gold, and I. S. Averbukh, J. Phys. Chem. Lett. **9**, 1105 (2018).
- [21] I. Tutunnikov, L. Xu, R. W. Field, K. A. Nelson, Y. Prior, and I. S. Averbukh, Phys. Rev. Res. **3**, 013249 (2021).
- [22] J. Bert, E. Prost, I. Tutunnikov, P. Béjot, E. Hertz, F. Billard, B. Lavorel, U. Steinitz, I. S. Averbukh, and O. Faucher, Laser Photon. Rev. **14**, 1900344 (2020).
- [23] I. Tutunnikov, E. Prost, U. Steinitz, P. Béjot, E. Hertz, F. Billard, O. Faucher, and I. S. Averbukh, Phys. Rev. A **104**, 053113 (2021).
- [24] B. Friedrich and D. R. Herschbach, Nature **353**, 412 (1991).
- [25] B. Friedrich and D. Herschbach, J. Chem. Phys. **111**, 6157 (1999).
- [26] L. Holmegaard, J. H. Nielsen, I. Nevo, H. Stapelfeldt, F. Filsinger, J. Küpper, and G. Meijer, Phys. Rev. Lett. **102**, 023001 (2009).
- [27] A. Goban, S. Minemoto, and H. Sakai, Phys. Rev. Lett. **101**, (2008).
- [28] O. Ghafur, A. Rouzée, A. Gijsbertsen, W. K. Siu, S. Stolte, and M. J. J. Vrakking, Nat. Phys. **5**, 289 (2009).
- [29] S. De, I. Znakovskaya, D. Ray, F. Anis, N. G. Johnson, I. A. Bocharova, M. Magrakvelidze, B. D. Esry, C. L. Cocke, I. V Litvinyuk, and M. F. Kling, Phys. Rev. Lett. **103**, (2009).
- [30] J. Hebling, K. Yeh, M. C. Hoffmann, B. Bartal, and K. A. Nelson, J. Opt. Soc. Am. B **25**, B6 (2008).
- [31] S. Fleischer, R. W. Field, and K. A. Nelson, Phys. Rev. Lett. **109**, 123603 (2012).
- [32] S. Fleischer, Y. Zhou, R. W. Field, and K. A. Nelson, Phys. Rev. Lett. **107**, 163603 (2011).
- [33] M. Shalaby and C. P. Hauri, Appl. Phys. Lett. **106**, 181108 (2015).
- [34] R. Damari, S. Kallush, and S. Fleischer, Phys. Rev. Lett. **117**, (2016).
- [35] R. Damari, D. Rosenberg, and S. Fleischer, Phys. Rev. Lett. **119**, 033002 (2017).

[36] A. Beer, R. Damari, Y. Chen, and S. Fleischer, J. Phys. Chem. A **126**, 3732 (2022).
[37] J. Dai, X. Xie, and X.-C. Zhang, Phys. Rev. Lett. **97**, 103903 (2006).
[38] E. L. Hahn, Phys. Rev. **80**, 580 (1950).
[39] R. G. Brewer and R. L. Shoemaker, Phys. Rev. Lett. **27**, 631 (1971).
[40] D. Zimdars, A. Tokmakoff, S. Chen, S. R. Greenfield, M. D. Fayer, T. I. Smith, and H. A. Schwettman, Phys. Rev. Lett. **70**, 2718 (1993).
[41] J. Hennig and R. K H I K Abt, Concepts Magn. Reson. **3**, 125 (1991).
[42] Y. Zhang, J. Shi, X. Li, S. L. Coy, R. W. Field, and K. A. Nelson, Proc. Natl. Acad. Sci. U. S. A. **118**, 2020941118 (2021).
[43] P. Wang, L. He, Y. He, S. Sun, R. Liu, B. Wang, P. Lan, and P. Lu, Opt. Express **29**, 663 (2021).